\documentclass[aps,prd,12pt,a4paper,groupedaddress,preprintnumbers,floatfix,nofootinbib,showpacs]{revtex4}
\usepackage[english]{babel}
\usepackage{amsmath}
\usepackage{graphicx}
\usepackage{subfig}
\usepackage{color}
\usepackage{float}
\usepackage{amssymb}
\usepackage{hyperref}
\usepackage{dcolumn}
\usepackage{bm}
\newcommand{\be}{\begin{equation}}
\newcommand{\ee}{\end{equation}}
\newcommand{\bea}{\begin{eqnarray}}
\newcommand{\eea}{\end{eqnarray}}

\def\lsim{\mathrel{\rlap{\lower4pt\hbox{\hskip1pt$\sim$}}
    \raise1pt\hbox{$<$}}}         
\def\gsim{\mathrel{\rlap{\lower4pt\hbox{\hskip1pt$\sim$}}
    \raise1pt\hbox{$>$}}}         

\begin{document}

\title{Relating quarks and leptons with the $T_{7}$ flavour group}

\author{
Cesar Bonilla,$^{1}$\footnote{Electronic address:cesar.bonilla@ific.uv.es}
Stefano Morisi,$^{2}$\footnote{Electronic address:stefano.morisi@gmail.com}
Eduardo Peinado,$^{3}$\footnote{Electronic address:epeinado@fisica.unam.mx}
and J. W. F. Valle,$^{1}$\footnote{Electronic address:valle@ific.uv.es}}

\affiliation{
  $^1$ Instituto de F\'{i}ısica Corpuscular (CSIC-Universitat de Val\`{e}ncia), Apdo. 22085, E-46071 Valencia, Spain. \\
  $^2$ DESY, Platanenallee 6, D-15735 Zeuthen, Germany. \\
  $^3$ Instituto de F\'isica, Universidad Nacional Aut\'onoma de M\'exico, 
  A.P. 20-364, M\'exico D.F. 01000, M\'exico.}  

\begin{abstract} 
  In this letter we present a model for quarks and leptons based on
  $T_{7}$ as flavour symmetry, predicting a canonical mass relation
  between charged leptons and down-type quarks proposed earlier.
  Neutrino masses are generated through a Type-I seesaw mechanism,
  with predicted correlations between the atmospheric mixing angle and
  neutrino masses.
  Compatibility with oscillation results lead to lower bounds for the
  lightest neutrino mass as well as for the neutrinoless double beta
  decay rates, even for normal neutrino mass hierarchy.
\end{abstract}
\pacs{11.30.Hv 14.60.-z 14.60.Pq 12.60.Fr 14.60.St 23.40.Bw}
\maketitle

\section{Introduction}

Ever since the discovery of the muon in the thirties particle
physicists have wondered on a possible simple understanding of fermion
mass and mixing patterns.
The experimental confirmation of neutrino oscillations
~\cite{Adamson:2011qu, An:2012eh, Ahn:2012nd, Abe:2011sj} has brought
again the issue into the spotlight.
Yet despite many attempts, so far the origin of neutrino mass and its
detailed flavour structure remains one of the most well-kept secrets
of nature.
In particular the observed values of neutrino oscillation
parameters~\cite{Forero:2014bxa} pose the challenge to figure out why
lepton mixing angles are so different to those of quarks.
Indeed the sharp differences between the flavour mixing parameters
characterizing the quark and lepton sectors escalate the complexity of
the flavour problem.
Many extensions of the Standard Model (SM) have been proposed in order
to induce nonzero neutrino masses~\cite{Schechter:1980gr} and to
predict the oscillation parameters such as the neutrino mass ordering,
the octant of the atmospheric mixing angle and the value of the
CP-violating phase in the lepton sector.

A popular approach to tackle these issues is the use of discrete
non-Abelian flavour symmetries which are known to be far more
restrictive than Abelian ones~\cite{Morisi:2012fg}. In the literature
there are many models based on, for instance, $A_{4}$ (the group of
even permutations of a tetrahedron) whose simplest realizations
predict zero reactor mixing angle and maximal atmospheric angle
\cite{Babu:2002dz,Ma:2004zv,Altarelli:2005yp}.  However, this nice
prediction has now been experimentally ruled out~\cite{Adamson:2011qu,
  An:2012eh, Ahn:2012nd, Abe:2011sj} so that the corresponding models
need to be revamped in order to account for
observations~\cite{Morisi:2013qna}.

A variety of possible predictions of flavour symmetry based models can
be found, for instance~\cite{King:2014nza}:
\begin{itemize}
  \item[i)]{\it neutrino mass sum rules} leading to restrictions on the
   effective mass parameter $|m_{ee}|$ characterizing neutrinoless
   double beta decay ($0\nu\beta\beta$) processes
   \cite{Barry:2010zk,Barry:2010yk,Dorame:2011eb,King:2013psa};
 \item[ii)] {\it neutrino mixing sum rules}~\cite{King:2005bj}.
\end{itemize}
Here we concentrate on the alternative possibility of having {\it mass
  relations} in the charged fermion sector. For definiteness we focus
on the relation in Eq.(\ref{MR2}),
\begin{equation}\label{MR2}
\frac{m_{b}}{\sqrt{m_{d}m_{s}}}\approx \frac{m_{\tau}}{\sqrt{m_{e}m_{\mu}}}.
\end{equation} 

This relation was suggested
in~\cite{Morisi:2011pt,Morisi:2013eca,Bazzocchi:2012ve,King:2013hj} and can hold at the
electroweak scale~\footnote{In an early paper~\cite{Wilczek:1978xi} Wilczek and Zee found,
  by using an $SU(2)_{H}$ symmetry, an extended mass relation
  $\frac{m_{b}}{\sqrt{m_{d}m_{s}}}=
  \frac{m_{\tau}}{\sqrt{m_{e}m_{\mu}}}=\frac{m_{t}}{\sqrt{m_{u}m_{c}}}$
  which is now evidently ruled out.}.
First we note that such a relation between down-type quark and charged
lepton masses can be understood because of group structure, when there
are three vacuum expectation values and only two invariant
contractions (Yukawas) in the product,
${\bf3}\otimes{\bf3}\otimes{\bf3}$.
For example, such relation can be obtained with other groups
containing three-dimensional irreducible representations (irreps) such
as, for example, $T_{n}\cong Z_{n}\rtimes Z_{3}$ (with
$n=7,13,19,31,43,49$; \cite{Ishimori:2010au}) as well as $T'$.

In what follows we build a flavour model for quarks and leptons based
upon the smallest non-Abelian group after $A_{4}$, namely the flavour
group
$T_{7}$~\cite{Luhn:2007sy,Luhn:2007yr,Cao:2010mp,Ishimori:2012sw,Kajiyama:2013lja,
  Vien:2014gza} leading to the mass relation in Eq.(\ref{MR2}).
Neutrino masses are generated by implementing a Type-I
seesaw~\cite{Valle:2006vb} in contrast to the dimensional-five
Weinberg operator approach used in previous
Refs.~\cite{Morisi:2011pt,Morisi:2013eca,King:2013hj}. We discuss the
resulting phenomenological predictions, namely, a correlation between
the lightest neutrino mass and the atmospheric angle, as well as lower
bounds for the effective mass parameter $|m_{ee}|$ characterizing
$0\nu\beta\beta$ decay for both neutrino mass orderings.

\section{The model}

Here we consider a model with the multiplet content in
Table~\ref{tmc1} where the SM electroweak gauge symmetry is crossed
with a global flavour symmetry group $T_{7}$. 
\begin{table}[!h]
\begin{tabular}{|c|c|c|c|c|c|c|c||c|c|c|c|c|c|}
\hline
                & $\overline{L}$& $\ell_{R}$& $N_{R}$& $\nu_{R}$&$\overline{Q}$& $d_{R}$ & $u_{R_{i}}$   & $H$     & $\varphi_{\nu}$& $\varphi_{u}$ & $\varphi_{d}$& $\xi_{\nu}$\\
\hline
$T_{7}$         &   ${\bf3}$    & ${\bf3}$ &${\bf3}$& ${\bf1}_{0}$  & ${\bf3}$     & ${\bf3}$&${\bf1}_{i}$&${\bf 1}_{0}$& $\bf3$         & $\bar{\bf3}$  & ${\bf3}$ & ${\bf1}_{0}$\\     
\hline
$\mathbb{Z}_{7}$&   $a^{3}$     &  $a^{3}$ & $a^{5}$& $a^{2}$ & $a^{3}$      & $a^{3}$ &  $a^{2}$   &  1      & $a^{4}$        &  $a^{2}$   &  $a^{1}$ &  $a^{3}$\\               
\hline
\end{tabular}\caption{Matter assignments of the model where $a^{7}=1$.}
\label{tmc1}
\end{table}
The down-type quarks and leptons (left- and right-handed) transform as
triplets, RH up-type quarks transform as singlets while the SM Higgs is
blind, as shown in Table~\ref{tmc1}. Then the Yukawa Lagrangian for
the charged sector is given by,
\begin{equation}\label{Lg1}
\mathcal{L}= \frac{Y^{\ell}}{\Lambda} \overline{L} \ell_{R} H_{d}+\frac{Y^{d}}{\Lambda} \overline{Q} d_{R} H_{d}+ \frac{Y^{u}}{\Lambda} \overline{Q} u_{R} H_{u}+h.c. 
\end{equation}
Here for simplicity we have omitted the flavour indices, and have
defined $H_{d}\equiv H\varphi_{d}$, $H_{u}\equiv \tilde{H}\varphi_{u}$
and $\tilde{H}\equiv i\sigma_{2}H^{\ast}$, where $\varphi_{a}$ are
$T_{7}$ flavon triplets and $\Lambda$ is the scale at which these
fields get their vacuum expectation values (vevs),
$\langle\varphi_{a}\rangle$.

On the other hand, let us assume the existence of four RH-neutrinos
accommodated as ${\bf3}\oplus{\bf1}_{0}$ under $T_{7}$ so that the
Lagrangian for the neutrino sector becomes,
\begin{equation}\label{Lnu1}
\mathcal{L}_{\nu}= \frac{Y^{\nu}_{1}}{\Lambda} \bar{L}N_{R} \tilde{H}_{d}+\frac{Y^{\nu}_{2}}{\Lambda} \bar{L}\nu_{R} H_{u}+
\kappa_{1} N_{R} N_{R} \varphi_{\nu}+ \kappa_{2} \nu_{R} \nu_{R} \xi_{\nu} 
\end{equation}
where, $\tilde{H}_{d}\equiv \tilde{H} \overline{\varphi}_{d}$. Notice
that the additional Abelian symmetry $\mathbb{Z}_{7}$ couples each
$T_{7}$ flavon triplet with only one fermion sector (down-type,
up-type or neutrino sector) , so that, flavons transform non-trivially
under the discrete Abelian group and their charges are unrelated to
each other by conjugation. Therefore, in some sense, the order of the
$\mathbb{Z}_{n}$ symmetry is fixed by the Yukawa sector.

In what follows we will study the flavon potential for three distinct
triplets under $T_{7}$.  The second column of Table~\ref{tal} shows
the vacuum expectation value alignments allowed in $T_{7}$
\cite{Luhn:2007sy,King:2013eh}, with small deviations from those
alignments shown in the third column.

\begin{table}[h]
\begin{tabular}{|c|c|c|}
\hline
\text{Flavon}    & VEV \text{Alignment} in $T_{7}$ &\text{Model}\\
\hline
 $\varphi_{\nu}$ & $(1,1,0)$  & $(1+\delta_{\nu_{1}},1,\delta_{\nu_{2}})$\\     
\hline
 $\varphi_{u}$   &  $(0,0,1)$  & $(\delta_{u_{1}},\delta_{u_{2}},1)$\\               
\hline
$\varphi_{d}$   &  $(1,0,0)$   & $(1,\delta_{d_{1}},\delta_{d_{2}})$\\               
\hline 
\end{tabular}\caption{Vacuum expectation value alignments.}
\label{tal}
\end{table}

\subsection{Flavon Potential}

The Higgs scalar potential for a single $T_{7}$ flavon triplet,
i.e. $\varphi\simeq{\bf3}$, is given by
\cite{Luhn:2007sy,King:2013eh}
\begin{equation}\label{fpot}
V_{s}=-\mu_{s}^2 \sum_{i=1}^{3}\varphi^{\dagger}_{i}\varphi_{i} +\lambda_{s}
\left(\sum_{i=1}^{3}\varphi^{\dagger}_{i}\varphi_{i}\right)^{2}+\kappa_{s} \sum_{i=1}^{3}\varphi^{\dagger}_{i}\varphi_{i}\varphi^{\dagger}_{i}\varphi_{i}.
\end{equation}
where the possible vacuum expectation value alignments are, see
Appendix~\ref{VAlign},
\begin{eqnarray}\label{als}
\langle\varphi\rangle\sim \frac{1}{\sqrt{3}}(1,1,1)\ \ \text{for}\ \ \kappa_{s}>0 \ \ \text{and}\ \ 
\langle\varphi\rangle\sim (1,0,0),(0,1,0),(0,0,1)\ \ \text{for}\ \ \kappa_{s}<0. 
\end{eqnarray}
In our case, ignoring the singlet $\xi_{\nu}$, there are three
triplets, $\varphi_u$, $\varphi_d$ and $\varphi_\nu$, with an
additional $\mathbb{Z}_{7}$ charge so that the flavon potential is
given as
\begin{equation}\label{Vt}
V'=V_{\nu}+V_{d}+V_{u}+V_{\text{mix}}, 
\end{equation}
where $V_{\alpha}$ (with $\alpha=\nu,d,u$) are given by
Eq.(\ref{fpot}). Then, in components, $V_{\alpha}$ contain the triplet
elements $\varphi_{\alpha_{i}}$ and the parameters $\mu_{\alpha}^{2}$,
$\lambda_{\alpha}$ and $\kappa_{\alpha}$.  The mixing part of the
potential is the following
\begin{equation}
 V_{\text{mix}}=\kappa_{12}\left|\sum_{i=1}^3\varphi_{\nu_{i}}^{\dagger}\varphi_{u_{i}}\right|^{2}+
 \kappa_{13}\left|\sum_{i=1}^3\varphi_{\nu_{i}}^{\dagger}\varphi_{d_{i}}\right|^{2}+
 \kappa_{23}\left|\sum_{i=1}^3\varphi_{d_{i}}^{\dagger}\varphi_{u_{i}}\right|^{2}+
 \kappa_{123}\left(\varphi_{\nu}\varphi_{u}\varphi_{d}+h.c.\right).
\end{equation}

The vev configuration written down in the second column of
Table~\ref{tal} is a minimum of the potential Eq.(\ref{Vt}) when
$\kappa_{\nu}>0$, $\kappa_{u}<0$ and $\kappa_{d}<0$ and the terms
$\kappa_{13}$ and $\kappa_{123}$, are suppressed.\footnote{The term
  proportional to $\kappa_{13}$ in the potential could be suppressed
  by adding a term like
  $-\mu_{13}^2(\varphi_{\nu}^{\dagger}\varphi_{d}+h.c.)$ which softly
  breaks $Z_{7}$. The trilinear term can be forbidden by invoking an
  additional parity transformation over the fields.}  Notice that some
vevs are orthogonal (namely,
$\langle\varphi_\nu\rangle\perp\langle\varphi_u\rangle$ and
$\langle\varphi_u\rangle\perp\langle\varphi_d\rangle$). This property
of the vevs has been described in \cite{King:2006np,King:2013eh}. 
In order to ensure a realistic model we assume small deviations
from the vev canonical alignments in the middle column in 
Table~\ref{tal}. Such deviations can be induced  by adding soft
breaking terms in the flavon potential, Eq.(\ref{Vt}).

\subsection{Mass relation in down-type sector}

As usual, one obtains the fermion mass matrices after electroweak
symmetry breaking from the Lagrangian in Eq.(\ref{Lg1}).
Given the $T_{7}$ multiplication rules (see Appendix~\ref{AppMR}),
one finds that the down--type quarks and the charged lepton mass
matrices turn out to have the form
\begin{eqnarray}
M_{f}=\left( 
\begin{array}{ccc}\label{Mf}
 0             & e^{i \theta_{f}} y_{1}^{f} v_{3}  & y_{2}^{f} v_{2} \\
 y_{2}^{f} v_{3}   & 0             & e^{i \theta_{f}} y_{1}^{f} v_{1}\\
e^{i \theta_{f}} y_{1}^{f} v_{2}  & y_{2}^{f} v_{1}   & 0
  \end{array}
  \right),
\end{eqnarray}
where $f=\ell,d$ and $\theta_{f}$ are unremovable phases contributing to 
CP-violation in the lepton and quark 
sector. In addition, we have used the following parameterization,
\begin{equation}\label{vdi}
 \frac{\langle\varphi_{d}\rangle \langle H \rangle}{\Lambda}\approx (v_{1},v_{2},v_{3}).
\end{equation}
It should be noticed that the matrices $M_{f}$ in Eq.(\ref{Mf}) have
the same structure as those obtained with $A_{4}$ as flavour
symmetry~\cite{Morisi:2009sc,Morisi:2011pt,Morisi:2013eca, King:2013hj}. It is
useful to rewrite Eq.(\ref{Mf}) in the following way,
\begin{eqnarray}
M_{f}=\left( 
\begin{array}{ccc}\label{Mlabr}
 0             & e^{i \theta_{f}} a^{f} \alpha^{f}  & b^{f} \\
 b^{f} \alpha^{f}   & 0             & e^{i \theta_{f}} a^{f} r^{f}\\
 e^{i \theta_{f}} a^{f}  & b^{f} r^{f}   & 0
  \end{array}
  \right),
\end{eqnarray}
where
\begin{eqnarray}\label{abalr}
a^{f}= y_{1}^{f} v_{2},\ \ b^{f}=y^{f}_{2} v_{2}, \ \ \alpha^{f}=v_{3}/v_{2}\ \ \text{and}\ \ r^{f}=v_{1}/v_{2}.
\end{eqnarray}
Following the reasoning in
\cite{Morisi:2011pt,Morisi:2013eca,King:2013hj} we consider the
invariants of $M_{f}M_{f}^{\dagger}$ and obtain, at leading order in
the limit $r^{f}>>\alpha^{f},1$ and $r^{f}>>b^{f}/a^{f}$,
\begin{eqnarray}
(b^{f} r^{f})^2 &\approx& m_3^2\label{br1},\\
b^{f\,6} r^{f\,2}\alpha^{f\,2} &\approx& m_1^2 m_2^2 m_3^2,\label{br2}\\
a^{f\,2} b^{f\,2} r^{f\,4}&\approx& m_2^2 m_3^2\label{aeq}.
\end{eqnarray}
Then, solving the last system of equations,
Eqs.(\ref{br1}-\ref{aeq}), one gets
\begin{eqnarray}\label{abr}
a^{f}\approx\frac{m_{2}}{m_{3}}\sqrt{\frac{m_{1}m_{2}}{\alpha^{f}}},\ \ b^{f}\approx\sqrt{\frac{m_{1}m_{2}}{\alpha^{f}}},\ \ \text{and}\ \
 r^{f}\approx m_{3}\sqrt{\frac{\alpha^{f}}{m_{1}m_{2}}}. 
\end{eqnarray}
From Eq.(\ref{abr}) and the fact that the same flavon is coupled to
the down-type quarks and charged leptons we are led to the mass
relation in Eq.(\ref{MR2}),
$$\frac{m_{b}}{\sqrt{m_{d}m_{s}}}\approx \frac{m_{\tau}}{\sqrt{m_{e}m_{\mu}}}.$$
It is worth mentioning that even when the phases $\theta_{f}$ appear
in the invariant $\text{det}|M_{f}M_{f}^{\dagger}|$ with $f=\ell,d$,
that is in Eq.(\ref{br2}), the mass relation is preserved.

\subsection{Quark mixing}

From the Yukawa Lagrangian in Eq.(\ref{Lg1}) we have that after
electroweak symmetry breaking the mass matrices for up- and down-type
quarks are, respectively,
\begin{eqnarray}\label{Mupd}
M_{u}&=\left(
\begin{array}{cccc}
 y^{u}_{1} u_{1} &            y^{u}_{2} u_{1} &           y^{u}_{3} u_{1}   \\
 y^{u}_{1} u_{2} & \omega     y^{u}_{2} u_{2} &\omega^{2} y^{u}_{3} u_{2}   \\
 y^{u}_{1} u_{3} & \omega^{2} y^{u}_{2} u_{3} &\omega     y^{u}_{3} u_{3}  
\end{array}
\right)\ \ \text{and}\ \
M_{d}=\left( 
\begin{array}{ccc}\label{Md}
 0             & e^{i \theta_{d}} a^{d} \alpha^{d}  & b^{d} \\
 b^{f} \alpha^{d}   & 0             & e^{i \theta_{d}} a^{d} r^{d}\\
 e^{i \theta_{d}} a^{d}  & b^{d} r^{d}   & 0
  \end{array}
  \right),
\end{eqnarray}
where the parameters $a^{d}$, $b^{d}$ and $r^{d}$ are given by
Eq.(\ref{abr}), with $\omega^{3}=1$ and the vevs $u_{i}$ ($i=1,2,3$)
defined through the parameterization
\begin{equation}\label{vui}
 \frac{\langle\varphi_{u}\rangle \langle H \rangle}{\Lambda}\approx (u_{1},u_{2},u_{3}),
\end{equation}
It is useful to rewrite the vevs as follows,
\begin{equation}\label{vu2}
(u_{1},u_{2},u_{3})=u_{3}\left(\frac{u_{1}}{u_{3}},\frac{u_{2}}{u_{3}},1\right)=u_{3}(\alpha_{1},\alpha_{2},1),
\end{equation}
in that way there are 10 free parameters in the quark sector, listed
in Table~\ref{tqp}. These parameters determine the six quark masses,
the three CKM mixing angles and the quark CP-violating phase.
\begin{table}[h!]
\begin{tabular}{|c|c|c|c|c|c|c|c|c|c|c|}
\hline
\text{10 free parameters} & $a^{d}$& $b^{d}$&$r^{d}$&$y_{1}^{u}$&$y_{2}^{u}$&$y_{3}^{u}$& $\alpha^{d}$ & $\alpha_{1}$ & $\alpha_{2}$& $\theta_d$\\
\hline
\end{tabular}\caption{Parameters characterizing the quark sector.}
\label{tqp}
\end{table}

In Ref.~\cite{King:2013hj} an $A_{4}$ flavour symmetry model was built
leading to our mass formula in Eq.~(\ref{MR2}). The mass and CKM mixing
parameters describing the quark sector, very similar to those in
Eq.(\ref{Mupd}), were successfully reproduced, as seen in in Table II
in \cite{King:2013hj}, assuming trivial phases, namely
$\theta_{d}=0,\pi$ in Eq.(\ref{Mupd}).
However, even in this trivial case there is CP-violation due to the
complex phase $\omega$. Here for simplicity we just take advantage of
the results given in \cite{King:2013hj} for the quark sector of our
current $T_{7}$ model.  Therefore we use the following values, given
in the aforementioned $A_{4}$ model,
\begin{eqnarray}\label{upvals}
&r^{d}=263.44 \text{MeV}, \ \ y_{1}^{u}u_{3}=-297393\, \text{MeV},\notag\\
&a^{d}=0.21\text{MeV}, \ \ y_{2}^{u}u_{3}=-15563\, \text{MeV}\notag\\
&b^{d}=10.73\text{MeV}, \ \ y_{3}^{u}u_{3}= 277\, \text{MeV}\notag\\
&\alpha^{d}=\frac{v_{3}}{v_{2}}=1.58,\ \ \alpha_{1}=\frac{u_{1}}{u_{3}}=2.14 \lambda^{4},\notag\\
&\theta_{d}=\pi, \ \ \text{and} \ \ \alpha_{2}=\frac{u_{2}}{u_{3}}=1.03 \lambda^{2},
 \end{eqnarray}
 and where $\lambda=0.2$ the Cabibbo angle. The parameters $r^{d}$,
 $a^{d}$ and $b^{d}$ can be computed by carrying out a substitution of
 $(m_{1},m_{2},m_{3})$ with the actual values of the down-type quark
 masses $(m_{d},m_{s},m_{b})$ in Eq.(\ref{abr}).  One can verify with
 ease that the predictions for the CKM mixing matrix, quark masses and
 CP-violation are in agreement with the experimental data
 \cite{Beringer:1900zz}.
 Now we proceed to study the lepton sector, for which some of the
 parameters will be fixed by the fit in the quark sector, namely the
 parameters $\alpha^{d}$ and $r^{d}$.

\subsection{Lepton mixing}

As we saw above, the spontaneous breaking of the electroweak symmetry
yields the following form for the charged lepton mass matrix,
\begin{eqnarray}
M_{\ell}=\left( 
\begin{array}{ccc}\label{Mlabr2}
 0             & e^{i \theta_{\ell}} a^{\ell} \alpha^{\ell}  & b^{\ell} \\
 b^{\ell} \alpha^{\ell}   & 0             & e^{i \theta_{\ell}} a^{\ell} r^{\ell}\\
 e^{i \theta_{\ell}} a^{\ell}  & b^{\ell} r^{\ell}   & 0
  \end{array}
  \right),
\end{eqnarray}
where, from the $T_{7}$ multiplication rules in the appendix one
finds,
\begin{eqnarray}\label{abrl}
a^{\ell}= y_{1}^{\ell} v_{2},\ \ b^{\ell}=y^{\ell}_{2} v_{2}, \ \ \alpha^{\ell}=v_{3}/v_{2}\ \ \text{and}\ \ r^{\ell}=v_{1}/v_{2}.
\end{eqnarray}

On the other hand, as mentioned in the introduction, here we adopt a
Type-I seesaw approach for generating the neutrino masses. This is in
contrast to previous models leading to the mass formula in
Eq.(\ref{MR2}) from the $A_{4}$ group.  In those schemes an effective
dimension-five operator approach was employed. In the present case the
neutrino mass matrix is given by,
\begin{eqnarray}\label{mnut1}
M_{\nu}=-M_{D}M_{RR}^{-1}M_{D}^{T}
\end{eqnarray}
where,
\begin{eqnarray}\label{MDRR}
M_{D}&=\left(
\begin{smallmatrix}
 Y^{\nu}_{1}v_{2} & 0 & 0 &e^{i \theta_{1}} Y^{\nu}_{2}u_{1}   \\
 0 &Y^{\nu}_{1} v_{3} & 0 &e^{i \theta_{1}} Y^{\nu}_{2}u_{2}   \\
 0 & 0 &Y^{\nu}_{1}v_{1}  &e^{i \theta_{1}} Y^{\nu}_{2}u_{3}  
\end{smallmatrix}
\right)\ \ \text{and}\ \
  M_{RR}=\left(
\begin{smallmatrix}
 0 & M_{3} & M_{2} & 0 \\
 M_{3} & 0 & M_{1} & 0 \\
 M_{2} & M_{1} & 0 & 0 \\
 0 & 0 & 0 & e^{i \theta_{2}}M_{4}
\end{smallmatrix}
\right),
\end{eqnarray}
where $M_{i}=\kappa_{1} \langle\varphi_{\nu}\rangle_{i}$ (for
$i=1,2,3$) and $M_{4}=\kappa_{2} \langle\xi_{\nu}\rangle$.  The real
matrix elements $M_{i}$ satisfy $M_{1}\sim M_{2}\gg M_{3}$, Table~\ref{tal}.
Notice that for complex Yukawas the mass matrices $M_{D}$
and $M_{RR}$ in Eq.(\ref{MDRR}) only depend on one unremovable phase.\\

In order to implement the vev alignments in Table~\ref{tal} we assume that the
vevs $u_{i}$ and $v_{i}$ in Eq.(\ref{MDRR}) satisfy $u_{3}\gg u_{1,2}$
and $v_{1}\gg v_{2,3}$.  The former vev hierarchy has to do with the
fit in the quark sector and the latter comes from the mass relation
$r^{d}\gg\alpha^{d},1$. Then, the vev alignments can be rewritten as
follows,
\begin{eqnarray}\label{vevsR}
\begin{array}{ccccc}
 u_{3}(\frac{u_{1}}{u_{3}},\frac{u_{2}}{u_{3}},1)&=&u(\alpha_{1},\alpha_{2},1)&\propto&(\delta_{u_{1}},\delta_{u_{2}},1),\\
 v_{2}(\frac{v_{1}}{v_{2}},1,\frac{v_{3}}{v_{2}})&=&v(r^{d},1,\alpha^{d})&\propto&(1,\delta_{d_{1}},\delta_{d_{2}}),\\
 M_{3}(\frac{M_{1}}{M_{3}},\frac{M_{2}}{M_{3}},1)&=& M(\epsilon_{1} R,R,1)&\propto& (1+\delta_{\nu_{1}},1,\delta_{\nu_{2}})
\end{array}
\end{eqnarray}
where $\alpha_{1}=2.14\lambda^{4}$, $\alpha_{2}=1.03\lambda^{2}$,
$\lambda=0.2$ and we have defined $u_{3}=u$, $v_{2}=v$ and $M_{3}=M$.

Therefore, using Eqs. (\ref{MDRR}-\ref{vevsR}), the light neutrino
mass matrix after the seesaw mechanism turns out to be
\begin{eqnarray}\label{Mnu2}
M_{\nu}&=\kappa
\left(
\begin{array}{ccc}
 \epsilon_{1}-2 e^{-i\theta_{\nu}} \alpha_{1}^2 \epsilon_{2} & -\alpha^{d}-2 e^{-i\theta_{\nu}} \alpha_{1} \alpha_{2} \epsilon_{2} & -\epsilon_{3}-2 e^{-i\theta_{\nu}} \alpha_{1} \epsilon_{2} \\
     \cdot   & \frac{\alpha^{d\, 2}}{\epsilon_{1}}-2 e^{-i\theta_{\nu}} \alpha_{2}^2 \epsilon_{2} & -\frac{\alpha^{d} \epsilon_{3}}{\epsilon_{1}}-2 e^{-i\theta_{\nu}} \alpha_{2} \epsilon_{2}\\
     \cdot   &    \cdot     & -2 e^{-i\theta_{\nu}} \epsilon_{2}+\frac{\epsilon_{3}^2}{\epsilon_{1}}
\end{array}
\right),
\end{eqnarray}
which is symmetric and $\alpha_{1}=2.14\lambda^{4}$,
$\alpha_{2}=1.03\lambda^{2}$, $\lambda=0.2$ and we have defined,

\begin{eqnarray}
 &\kappa\equiv \frac{(Y^{\nu}v)^{2}}{M}, \ \ \epsilon_{2}\equiv \frac{M (Y^{\nu}_{2} u)^{2}}{M_{4} (Y^{\nu}_{1} v)^{2}},\ \ \epsilon_{3}\equiv \frac{r^{d}}{R}\notag\\
 &\text{and} \ \ \theta_{\nu}\equiv -2 \theta_{1}+\theta_{2}.
\end{eqnarray}

It is important to note that some parameters in the neutrino mass
matrix are fixed by the fit in the quark sector.  In Table~\ref{tlp}
we list the parameters in the lepton sector denoting as ``fixed''
those determined by the fit in the quark sector.  Bear in mind that
down-type quarks and charged leptons couple to the same flavon
$\varphi_{d}$ and hence, $\alpha^{d}=\alpha^{\ell}$ and
$r^{d}=r^{\ell}$. This is the origin of the mass relation in
Eq.~(\ref{MR2}).
\begin{table}
\begin{tabular}{|c|c|c|c|c|c|c|c|c|c|c|c|}
\hline
 \text{Parameters in the lepton sector}  & $a^{\ell}$ & $b^{\ell}$& $r^{d}$ &$\alpha^{d}$&$\alpha_{1}$&$\alpha_{2}$&$\epsilon_{1}$& $\epsilon_{2}$ & $\epsilon_{3}$ & $\theta_{\ell}$ & $\theta_{\nu}$\\
\hline
\text{Fixed}                             &            &           &\checkmark&\checkmark &\checkmark &  \checkmark&    &      &              &                 &   \\                                
\hline
\text{Free}                              & \checkmark & \checkmark &          &           &          &           &\checkmark &  \checkmark      & \checkmark    & \checkmark    & \checkmark   \\ 
\hline                                  
\end{tabular}\caption{Parameters in the lepton sector.}
\label{tlp}
\end{table}

Gathering all we have in the lepton sector we can compute the lepton
mixing matrix,
\begin{equation}
U=U_{\ell}^{\dagger}U_{\nu} 
\end{equation}
where $U_{\ell}$ and $U_{\nu}$ are the matrices that diagonalize the
charged and neutral mass matrices, $M_{\ell}^{2}\equiv
M_{\ell}M_{\ell}^{\dagger}$ and $M_{\nu}^{2}\equiv
M_{\nu}M_{\nu}^{\dagger}$, respectively. Remind that $M_{\ell}$ is the
matrix in Eq.(\ref{Mlabr2}) with one unremovable phase
$\theta_{\ell}$.

\section{Results}

In our analysis, we have varied for instance $\epsilon_{i}$ in the
range $\left[0,5\right]$ and the phases $\theta_{\ell,\nu}$ in the
range $\left[0,2\pi\right]$. We make use of the neutrino mass matrix
invariants $\text{tr}M_{\nu}^{2}$, $\text{det}M_{\nu}^{2}$ and
$(\text{tr}M^{2}_{\nu})^2-\text{tr}(M_{\nu}^4)$ and choose to rewrite
the three neutrino masses in terms of the square mass differences
$\Delta m_{\text{atm}}^{2}$ and $\Delta m_{\text{sol}}^{2}$ and the
lightest neutrino mass, $m_{1}$ for the case of normal hierarchy and
$m_{3}$ for inverted hierarchy.  We now sum up all our results.\\

\begin{figure}[H]
  \centering
  {\label{}\includegraphics[width=0.45\textwidth]{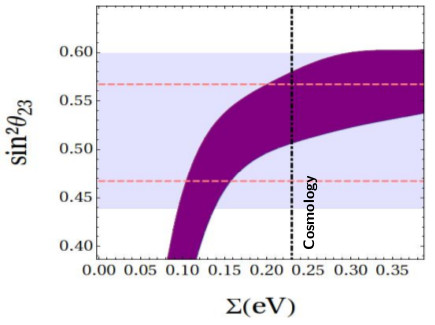}}\ \
  {\label{}\includegraphics[width=0.45\textwidth]{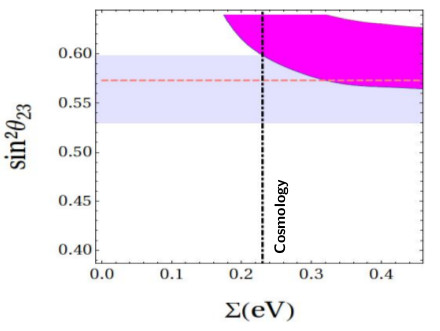}}\\
  \caption{Left panel: Correlation between the atmospheric angle and
    the sum of neutrino masses $\Sigma$ for the normal hierarchy case.
    Right panel: Correlation between the atmospheric angle and
    $\Sigma$ when assuming inverted hierarchy.  The horizontal dotted
    lines denote the best fit values for the atmospheric
    angle~\cite{Forero:2014bxa} while the horizontal bands are allowed
    at $1\sigma$. The vertical dot-dashed line is the cosmological
    bound from the combination of CMB and BAO data~\cite{Ade:2013zuv}}
  \label{fig:1}
\end{figure}
\begin{figure}[H]
      \centering
                \includegraphics[width=12cm]{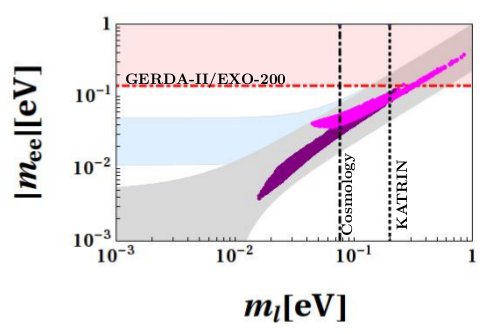}
                \caption{Effective neutrino mass parameter $|m_{ee}|$
                  versus the lightest neutrino mass for normal
                  (purple/dark region) and inverted (magenta/light
                  region) hierarchies. The vertical dotdashed line and
                  labeled as ``Cosmology'' denotes the bound from the
                  combination of CMB and BAO
                  data~\cite{Ade:2013zuv}. The vertical dotted line is
                  the future sensitivity of KATRIN,
                  \cite{Bornschein:2003xi}.  Here the oscillation
                  constraints are taken at
                  $3\sigma$~\cite{Forero:2014bxa}.}
         \label{meeml}
         \end{figure}

The panel on the left in Fig.\ref{fig:1} shows the correlation between
the atmospheric angle for normal hierarchy (NH, i.e.
$|m_{3}|>|m_{2}|>|m_{1}|$) and the sum of neutrino masses (defined as
$\Sigma\equiv|m_{1}|+|m_{2}|+|m_{3}|$). We find that there is a lower
bound for the lightest neutrino mass and that the first octant is
favored by lighter neutrino masses.  For reference we also display the
constraint coming from the combination of cosmological CMB data from
Planck and WMAP, including baryon acoustic oscillations (BAO) data
from~\cite{Ade:2013zuv}.
If taken at face value such stringent cosmological bound would
disfavor not only heavy neutrinos but also the best fit value for the
atmospheric angle lying in second octant \cite{Forero:2014bxa}.
On the other hand, a similar correlation between the atmospheric angle
and the sum of neutrino masses, $\Sigma$, is also found for the
inverted hierarchy case (IH, i.e. $|m_{2}|>|m_{1}|>|m_{3}|$).
This is shown on the right panel of Fig.\ref{fig:1} where the
dot-dashed vertical line is the constraint coming from the same
combination of cosmological data \cite{Ade:2013zuv}.  Taking the most
stringent cosmological (BAO) bound into account as well as the
oscillation results one sees that, at $1\sigma$, this case would be
disfavored.
Indeed, if this cosmological bound is taken at face value, the second
octant would be excluded for inverse hierarchy. However, as seen in
Fig.~\ref{meeml}, at $3\sigma$ the second octant is certainly allowed
for inverted hierarchy.  The resulting lower bound for the lightest
neutrino mass is much tighter than the one that holds for normal
hierarchy.
For comparison we also display the future sensitivity of the KATRIN
experiment on tritium beta decay, $\Sigma\simeq 0.6$ eV,
\cite{Bornschein:2003xi}.

In summary, one sees that for both hierarchies our model implies a
correlation between the atmospheric angle and the lightest neutrino
mass. The current neutrino oscillation experiments lead to a lower
bound for $m_{1}$.

Such a lower bounds have implications for the effective mass parameter
$|m_{ee}|$ specifying the neutrinoless double beta -- $0\nu\beta\beta$
-- decay amplitude.

Let us now turn to the implications for $0\nu\beta\beta$.
In Fig. (\ref{meeml}) we plot the effective parameter $|m_{ee}|$ as
function of the lightest neutrino mass.  The NH case corresponds to
the purple/dark region, while the IH case is denoted by the
magenta/light region, respectively. The vertical dot-dashed line and
labeled as ``Cosmology'' represents the constraint coming from the
combination of CMB data~\cite{Ade:2013zuv}, as well as the future
sensitivity of KATRIN~\cite{Bornschein:2003xi} indicated by the
vertical dotted line.  

\section{Conclusions}

In this paper we have suggested a model based on the flavour symmetry
group $T_{7}$ leading to a very successful canonical mass relation
between charged leptons and down-type quarks proposed
in~\cite{Morisi:2011pt,Morisi:2013eca,King:2013hj}.
Previous papers predicting this mass relation have adopted the $A_{4}$
flavour symmetry and assumed that neutrino masses were generated
through higher order operators.
In our $T_{7}$ model the neutrino masses are generated through the
conventional Type-I seesaw mechanism. 

The model leads to a correlation between the lightest neutrino mass
and the atmospheric angle. This correlation implies lower bounds for
the lightest neutrino mass which come from applying the neutrino
oscillation constraints. These bounds on the lightest neutrino mass
also translate to lower bounds on the effective amplitude parameter
$|m_{ee}|$ characterizing $0\nu\beta\beta$ decay for both neutrino
mass hierarchies.

\section{Acknowledgments}

  Work supported by the Spanish grants FPA2011-22975 and Multidark
  CSD2009-00064 (MINECO), and PROMETEOII/2014/084 (Generalitat
  Valenciana).  

\appendix
\section{Vacuum Alignments}\label{VAlign}
  Let us assume that the vev of the $T_7$ flavon triplet is 
  real and that the field is shifted as, 
  \be 
  \varphi_i=u_i+\phi_i.  
  \ee 
  The flavon potential is given by~\cite{Luhn:2007sy},
\begin{equation}\label{fpot}
V_{s}=-\mu_{s}^2 \sum_{i=1}^{3}\varphi^{\dagger}_{i}\varphi_{i} +\lambda_{s}
\left(\sum_{i=1}^{3}\varphi^{\dagger}_{i}\varphi_{i}\right)^{2}+\kappa_{s} 
\sum_{i=1}^{3}\varphi^{\dagger}_{i}\varphi_{i}\varphi^{\dagger}_{i}\varphi_{i},
\end{equation}
where $\lambda_s>0$ .
The minimization conditions are obtained by taking,
\be
\left.\frac{\partial V_{s}}{\partial \varphi_i}\right|_{\varphi_{i}\to0}=0,
\ee
which leads to the following system of equations,
\bea\label{seq}
&&-\mu^2+2(\kappa_s +\lambda_s ) u_1^2+ 2\lambda_s (u_2^2+u_3^2)=0\notag\\
&&-\mu^2+2(\kappa_s +\lambda_s ) u_2^2+ 2\lambda_s (u_1^2+u_3^2)=0 \\
&&-\mu^2+2(\kappa_s +\lambda_s ) u_3^2+ 2\lambda_s (u_1^2+u_2^2)=0.\notag
\eea
One set of minimization conditions is obtained by solving~(\ref{seq}) 
for instance for $\mu^2$, $u_2$ and $u_3$ ,
\bea\label{mincon}
a)&&\mu^2=2(\kappa_s+3\lambda_s) u_1^2,\ \ u_2=u_3=u_1; \notag\\  
b)&&\mu^2=2(\kappa_s+\lambda_s) u_1^2,\ \ u_2=u_3=0; \\
c)&&\mu^2=2(\kappa_s+2\lambda_s) u_1^2,\ \ u_2=u_1\ \ \text{and}\ \ u_3=0,  \notag
\eea
which can be translated in the following alignments, $\langle \varphi 
\rangle\equiv (u_1 ,u_2, u_{3})\sim(1,1,1)$, $ \langle \varphi 
\rangle\sim (1,0,0)$ and $\langle \varphi\rangle\sim(1,1,0)$, 
respectively.
In order to characterize each case in~(\ref{mincon}) as a local
minimum we compute the Hessian matrix, \be
\left.\mathcal{H}=\frac{\partial^2 V_{s}}{\partial \varphi_i
\partial \varphi_j}\right|_{\varphi_{i}\to0},
\ee
and verify its positivity, that is all its eigenvalues
are positive. For case $a)$ the Hessian matrix turns out to be,
\bea\label{Hessa}
\mathcal{H}_{a}=
8 u_1^2\left(
\begin{array}{ccc}
 (\kappa_s +\lambda_s ) &  \lambda_s &  \lambda_s   \\
 \lambda_s   &  (\kappa_s +\lambda_s ) &  \lambda_s  \\
 \lambda_s   &  \lambda_s  &  (\kappa_s +\lambda_s ) 
\end{array}
\right).
\eea
The eigenvalues of $\mathcal{H}_{a}$ are, $8 u_1^2 (\kappa_s, \kappa_s , 
\kappa_s +3 \lambda_s )$ 
which are positive iff $\kappa_s>0$. For $b)$ we have, 
\bea\label{Hessb}
\mathcal{H}_{b}=
4u_1^2\left(
\begin{array}{ccc}
 2 (\kappa_s +\lambda_s ) & 0 & 0 \\
 0 & - \kappa_s  & 0 \\
 0 & 0 & - \kappa_s 
\end{array}
\right),
\eea
which is positive definite if $-\lambda_s<\kappa_s<0$. Finally, in the last
case we have,
\bea\label{Hessc}
\mathcal{H}_{c}=
4u_1^2\left(
\begin{array}{ccc}
 2 (\kappa_s +\lambda_s ) & 2 \lambda_s   & 0 \\
 2 \lambda_s   & 2 (\kappa_s +\lambda_s ) & 0 \\
 0 & 0 & - \kappa_s  
\end{array}
\right).
\eea

The eigenvalues of $\mathcal{H}_{c}$ are given by, $4u_1^2(2 \kappa_s
,2 (\kappa_s +2 \lambda_s ), - \kappa_s )$. 
Therefore, we have that the only possible {\it global} minima are,
\bea
a)&&\langle \varphi\rangle\sim (\pm1,\pm1,\pm1)\ \ \text{for}\ \ \kappa_s >0,\notag\\  
b)&&\langle \varphi\rangle\sim (\pm1,0,0) \ \ \text{for}\ \ -\lambda_s<\kappa_s <0\notag
\eea
up to sign permutations in the former and permutations of the
non-zero value in the latter. These other possibilities lead to
degenerate vacua. In the realistic case of our model there
are other terms in the potential including $T_7$ symmetry breaking
terms needed to generate $\delta$s in Table~\ref{tal}. In general
these are expected to lift the degeneracies of the above minima.

\section{$T_{7}$ group basics}\label{AppMR}

The group $T_{7}$ is a subgroup of $SU(3)$ with 21 elements and
isomorphic to $\mathbb{Z}_{7}\rtimes\mathbb{Z}_{3}$.  This group has
five irreducible representations (e.i., ${\bf 1}_{0}$, ${\bf1}_{1}$,
${\bf1}_{2}$, ${\bf 3}$ and $\bar{\bf 3}$) and is known as the
smallest group containing a complex triplet.  The multiplication rules
in $T_{7}$ are the following,
\begin{eqnarray}
 \bf{3}\otimes{\bf3}=\bf{3}\oplus\bar{\bf3}\oplus\bar{\bf3},\ \ \bf{3}\otimes{\bf3}=\bar{\bf{3}}\oplus\bf3\oplus{\bf3},\notag\\
 \bf{3}\otimes\bar{\bf 3}=\sum_{a=0}^{2}{\bf1}_{a}\oplus{\bf3}\oplus\bar{\bf3}\ \ \text{and}\ \ \bf{3}\otimes{\bf1}=\bf{3}.
\end{eqnarray}
Let ${\bf X}^{a}=(x^{a}_{1},x^{a}_{2},x^{a}_{3})^{T}$, $\bar{\bf X}^{a}=(\bar{x}^{a}_{1},\bar{x}^{a}_{2},\bar{x}^{a}_{3})^{T}$,
and ${\bf z}_{i}$ (with $i=0,1,2$), be triplets, anti-triplets and singlets, respectively, under $T_{7}$ then 
these elements are multiplied as follows:
\begin{eqnarray}
\bullet &{\bf X} \times {\bf X}'={\bf X}''+\bar{\bf X}+\bar{\bf X}',\ \ \text{where}\ \ 
{\bf X}''=(x_{3}x'_{3}, x_{1}x'_{1},x_{2}x'_{2}), 
\bar{\bf X}=(x_{2} x'_{3},x_{3} x'_{1},x_{1}x'_{2})\notag\\
&\text{and} \ \  \bar{\bf X}'=(x_{3}x'_{2},x_{1}x'_{3}, x_{2}x'_{1}),\\
\bullet &{\bf X}\times \bar{\bf X}= \sum_{a=0}^{2}{\bf z}_{a}+{\bf X}'+ \bar{\bf X}',\ \ \text{where}\ \ 
z_{a}=x_{1}\bar{x}_{1}+\omega^{2a}x_{2}\bar{x}_{2}+\omega^{a} x_{3}\bar{x}_{3},\notag\\
&{\bf X}'=(x_{2}\bar{x}_{1},x_{3}\bar{x}_{2},x_{1}\bar{x}_{3}),\ \
\text{and}\ \ \bar{\bf X}'=(x_{1}\bar{x}_{2},x_{2}\bar{x}_{3},x_{3}\bar{x}_{1}),\\
\bullet&{\bf z}_{a}\times{\bf X} = {\bf X}'\ \ \text{where}\ \ {\bf X}'=(z_{a} x_{1},\omega^{a} z_{a} x_{2},\omega^{2a} z_{a} x_{3}).
\end{eqnarray}

For more details about the group $T_{7}$ see for instance,
Refs.~\cite{Luhn:2007sy,Luhn:2007yr,Ishimori:2010au}.


\begin{thebibliography}{10}

\bibitem{Adamson:2011qu}
{\bfseries MINOS Collaboration} Collaboration, P.~Adamson {\em et~al.},
  ``{Improved search for muon-neutrino to electron-neutrino oscillations in
  MINOS},'' {\em Phys.Rev.Lett.} {\bfseries 107} (2011) 181802,
\href{http://arxiv.org/abs/1108.0015}{{\ttfamily arXiv:1108.0015 [hep-ex]}}.

\bibitem{An:2012eh}
{\bfseries DAYA-BAY Collaboration} Collaboration, F.~An {\em et~al.},
  ``{Observation of electron-antineutrino disappearance at Daya Bay},'' {\em
  Phys.Rev.Lett.} {\bfseries 108} (2012) 171803,
\href{http://arxiv.org/abs/1203.1669}{{\ttfamily arXiv:1203.1669 [hep-ex]}}.

\bibitem{Ahn:2012nd}
{\bfseries RENO collaboration} Collaboration, J.~Ahn {\em et~al.},
  ``{Observation of Reactor Electron Antineutrino Disappearance in the RENO
  Experiment},'' {\em Phys.Rev.Lett.} {\bfseries 108} (2012) 191802,
\href{http://arxiv.org/abs/1204.0626}{{\ttfamily arXiv:1204.0626 [hep-ex]}}.

\bibitem{Abe:2011sj}
{\bfseries T2K Collaboration} Collaboration, K.~Abe {\em et~al.}, ``{Indication
  of Electron Neutrino Appearance from an Accelerator-produced Off-axis Muon
  Neutrino Beam},'' {\em Phys.Rev.Lett.} {\bfseries 107} (2011) 041801,
  \href{http://arxiv.org/abs/1106.2822}{{\ttfamily arXiv:1106.2822 [hep-ex]}}.

\bibitem{Forero:2014bxa}
D.~Forero, M.~Tortola, and J.~Valle, ``{Neutrino oscillations refitted},''
\href{http://dx.doi.org/10.1103/PhysRevD.90.093006}
{{\em Phys. Rev.} {\bfseries D90}, 093006 (2014)},
\href{http://arxiv.org/abs/1405.7540}{{\ttfamily arXiv:1405.7540 [hep-ph]}},
 further references in
``Global status of neutrino oscillation parameters after Neutrino-2012,''
  Phys.\ Rev.\ D {\bf 86}, 073012 (2012)

 \bibitem{Schechter:1980gr}
J.~Schechter and J.~Valle, ``{Neutrino Masses in SU(2) x U(1) Theories},''
\href{http://dx.doi.org/10.1103/PhysRevD.22.2227}{{\em Phys.Rev.} {\bfseries
  D22} (1980) 2227}.

\bibitem{Morisi:2012fg}
S.~Morisi and J.~Valle, ``{Neutrino masses and mixing: a flavour symmetry
  roadmap},'' {\em Fortsch.Phys.} {\bfseries 61} (2013) 466--492,
\href{http://arxiv.org/abs/1206.6678}{{\ttfamily arXiv:1206.6678 [hep-ph]}}.

\bibitem{Babu:2002dz}
K.~Babu, E.~Ma, and J.~Valle, ``{Underlying A(4) symmetry for the neutrino mass
  matrix and the quark mixing matrix},''
  \href{http://dx.doi.org/10.1016/S0370-2693(02)03153-2}{{\em Phys.Lett.}
  {\bfseries B552} (2003) 207--213},
  \href{http://arxiv.org/abs/hep-ph/0206292}{{\ttfamily arXiv:hep-ph/0206292
  [hep-ph]}}.

\bibitem{Ma:2004zv}
E.~Ma, ``A(4) origin of the neutrino mass matrix,'' {\em Phys. Rev.} {\bfseries
  D70} (2004) 031901,
\href{http://arxiv.org/abs/hep-ph/0404199}{{\ttfamily hep-ph/0404199}}.

\bibitem{Altarelli:2005yp}
G.~Altarelli and F.~Feruglio, ``Tri-bimaximal neutrino mixing from discrete
  symmetry in extra dimensions,'' {\em Nucl. Phys.} {\bfseries B720} (2005)
  64--88,
\href{http://arxiv.org/abs/hep-ph/0504165}{{\ttfamily hep-ph/0504165}}.

\bibitem{Morisi:2013qna}
S.~Morisi, D.~Forero, J.~Romao, and J.~Valle, ``{Neutrino mixing with revamped
  A4 flavour symmetry},'' {\em Phys.Rev.} {\bfseries D88} (2013) 016003,
\href{http://arxiv.org/abs/1305.6774}{{\ttfamily arXiv:1305.6774 [hep-ph]}}.

\bibitem{King:2014nza}
S.~F. King, A.~Merle, S.~Morisi, Y.~Shimizu, and M.~Tanimoto, ``{Neutrino Mass
  and Mixing: from Theory to Experiment},''
  \href{http://dx.doi.org/10.1088/1367-2630/16/4/045018}{{\em New J.Phys.}
  {\bfseries 16} (2014) 045018},
\href{http://arxiv.org/abs/1402.4271}{{\ttfamily arXiv:1402.4271 [hep-ph]}}.

\bibitem{Barry:2010zk}
J.~Barry and W.~Rodejohann, ``{Deviations from tribimaximal mixing due to the
  vacuum expectation value misalignment in A4 models},''
  \href{http://dx.doi.org/10.1103/PhysRevD.81.119901,
  10.1103/PhysRevD.81.093002}{{\em Phys.Rev.} {\bfseries D81} (2010) 093002},
\href{http://arxiv.org/abs/1003.2385}{{\ttfamily arXiv:1003.2385 [hep-ph]}}.

\bibitem{Barry:2010yk} 
  J.~Barry and W.~Rodejohann,``{Neutrino Mass Sum-rules in Flavor Symmetry 
  Models},''
  \href{http://dx.doi.org/10.1016/j.nuclphysb.2010.08.015} 
  {{\em Nucl.Phys.} {\bfseries B842}, 33 (2011)},
  \href{http://arxiv.org/abs/1007.5217}{{\ttfamily arXiv:1007.5217 [hep-ph]}}.

\bibitem{Dorame:2011eb}
L.~Dorame, D.~Meloni, S.~Morisi, E.~Peinado, and J.~Valle, ``{Constraining
  Neutrinoless Double Beta Decay},'' {\em Nucl.Phys.} {\bfseries B861} (2012)
  259--270,
\href{http://arxiv.org/abs/1111.5614}{{\ttfamily arXiv:1111.5614 [hep-ph]}}.

\bibitem{King:2013psa}
S.~F. King, A.~Merle, and A.~J. Stuart, ``{The Power of Neutrino Mass Sum Rules
  for Neutrinoless Double Beta Decay Experiments},''
  \href{http://dx.doi.org/10.1007/JHEP12(2013)005}{{\em JHEP} {\bfseries 1312}
  (2013) 005},
\href{http://arxiv.org/abs/1307.2901}{{\ttfamily arXiv:1307.2901 [hep-ph]}}.

\bibitem{King:2005bj}
S.~King, ``{Predicting neutrino parameters from SO(3) family symmetry and
  quark-lepton unification},''
  \href{http://dx.doi.org/10.1088/1126-6708/2005/08/105}{{\em JHEP} {\bfseries
  0508} (2005) 105},
\href{http://arxiv.org/abs/hep-ph/0506297}{{\ttfamily arXiv:hep-ph/0506297
  [hep-ph]}}.

\bibitem{Morisi:2011pt}
S.~Morisi, E.~Peinado, Y.~Shimizu, and J.~Valle, ``{Relating quarks and leptons
  without grand-unification},'' {\em Phys.Rev.} {\bfseries D84} (2011) 036003,
\href{http://arxiv.org/abs/1104.1633}{{\ttfamily arXiv:1104.1633 [hep-ph]}}.

\bibitem{Morisi:2013eca}
S.~Morisi, M.~Nebot, K.~M. Patel, E.~Peinado, and J.~Valle, ``{Quark-Lepton
  Mass Relation and CKM mixing in an A4 Extension of the Minimal Supersymmetric
  Standard Model},'' \href{http://dx.doi.org/10.1103/PhysRevD.88.036001}{{\em
  Phys.Rev.} {\bfseries D88} (2013) 036001},
\href{http://arxiv.org/abs/1303.4394}{{\ttfamily arXiv:1303.4394 [hep-ph]}}.

\bibitem{King:2013hj}
S.~King, S.~Morisi, E.~Peinado, and J.~Valle, ``{Quark-Lepton Mass Relation in
  a Realistic A4 Extension of the Standard Model},'' {\em Phys. Lett. B}
  {\bfseries 724} (2013) 68--72,
\href{http://arxiv.org/abs/1301.7065}{{\ttfamily arXiv:1301.7065 [hep-ph]}}.

\bibitem{Bazzocchi:2012ve} 
  F.~Bazzocchi, S.~Morisi, E.~Peinado, J.~W.~F.~Valle and A.~Vicente, 
  ``{Bilinear R-parity violation with flavor symmetry},'' 
  \href{http://dx.doi.org/10.1007/JHEP01(2013)033}
  {{\em JHEP} {\bfseries 1301}, 033 (2013)}
  \href{http://arxiv.org/abs/1202.1529}{{\ttfamily arXiv:1202.1529 [hep-ph]}}.

\bibitem{Wilczek:1978xi} 
  F.~Wilczek and A.~Zee,
  ``{Horizontal Interaction and Weak Mixing Angles},''
  \href{http://dx.doi.org/10.1103/PhysRevLett.42.421}
  {{\em Phys. Rev. Lett} {\bfseries 42}, 421 (1979)}.
 
\bibitem{Ishimori:2010au}
H.~Ishimori, T.~Kobayashi, H.~Ohki, Y.~Shimizu, H.~Okada, {\em et~al.},
  ``{Non-Abelian Discrete Symmetries in Particle Physics},''
  \href{http://dx.doi.org/10.1143/PTPS.183.1}{{\em Prog.Theor.Phys.Suppl.}
  {\bfseries 183} (2010) 1--163},
\href{http://arxiv.org/abs/1003.3552}{{\ttfamily arXiv:1003.3552 [hep-th]}}.

\bibitem{Luhn:2007sy}
C.~Luhn, S.~Nasri, and P.~Ramond, ``{Tri-bimaximal neutrino mixing and the
  family symmetry semidirect product of Z(7) and Z(3)},''
  \href{http://dx.doi.org/10.1016/j.physletb.2007.06.059}{{\em Phys.Lett.}
  {\bfseries B652} (2007) 27--33},
\href{http://arxiv.org/abs/0706.2341}{{\ttfamily arXiv:0706.2341 [hep-ph]}}.

\bibitem{Luhn:2007yr}
C.~Luhn, S.~Nasri, and P.~Ramond, ``{Simple Finite Non-Abelian Flavor
  Groups},'' \href{http://dx.doi.org/10.1063/1.2823978}{{\em J.Math.Phys.}
  {\bfseries 48} (2007) 123519},
\href{http://arxiv.org/abs/0709.1447}{{\ttfamily arXiv:0709.1447 [hep-th]}}.

\bibitem{Cao:2010mp}
Q.-H. Cao, S.~Khalil, E.~Ma, and H.~Okada, ``{Observable T7 Lepton Flavor
  Symmetry at the Large Hadron Collider},''
  \href{http://dx.doi.org/10.1103/PhysRevLett.106.131801}{{\em Phys.Rev.Lett.}
  {\bfseries 106} (2011) 131801},
\href{http://arxiv.org/abs/1009.5415}{{\ttfamily arXiv:1009.5415 [hep-ph]}}.

\bibitem{Ishimori:2012sw}
H.~Ishimori, S.~Khalil, and E.~Ma, ``{CP Phases of Neutrino Mixing in a
  Supersymmetric B-L Gauge Model with T7 Lepton Flavor Symmetry},''
  \href{http://dx.doi.org/10.1103/PhysRevD.86.013008}{{\em Phys.Rev.}
  {\bfseries D86} (2012) 013008},
\href{http://arxiv.org/abs/1204.2705}{{\ttfamily arXiv:1204.2705 [hep-ph]}}.

\bibitem{Kajiyama:2013lja}
Y.~Kajiyama, H.~Okada, and K.~Yagyu, ``{T7 Flavor Model in Three Loop Seesaw
  and Higgs Phenomenology},''
  \href{http://dx.doi.org/10.1007/JHEP10(2013)196}{{\em JHEP} {\bfseries 1310}
  (2013) 196},
\href{http://arxiv.org/abs/1307.0480}{{\ttfamily arXiv:1307.0480 [hep-ph]}}.

\bibitem{Vien:2014gza}
V.~Vien and H.~Long, ``{The T7 flavor symmetry in 3-3-1 model with neutral
  leptons},'' \href{http://dx.doi.org/10.1007/JHEP04(2014)133}{{\em JHEP}
  {\bfseries 1404} (2014) 133},
\href{http://arxiv.org/abs/1402.1256}{{\ttfamily arXiv:1402.1256 [hep-ph]}}.

\bibitem{Valle:2006vb}
J.~W.~F. Valle, ``Neutrino physics overview,'' {\em J. Phys. Conf. Ser.}
  {\bfseries 53} (2006) 473--505,
  \href{http://arxiv.org/abs/hep-ph/0608101}{{\ttfamily hep-ph/0608101}}.
These review lectures were given at Corfu, 2005 and contain extensive
  references to the early papers on the seesaw mechanism.

\bibitem{King:2013eh}
S.~F. King and C.~Luhn, ``{Neutrino Mass and Mixing with Discrete Symmetry},''
  {\em Rept.Prog.Phys.} {\bfseries 76} (2013) 056201,
\href{http://arxiv.org/abs/1301.1340}{{\ttfamily arXiv:1301.1340 [hep-ph]}}.

\bibitem{King:2006np}
S.~F. King and M.~Malinsky, ``A(4) family symmetry and quark-lepton
  unification,'' {\em Phys. Lett.} {\bfseries B645} (2007) 351--357,
\href{http://arxiv.org/abs/hep-ph/0610250}{{\ttfamily hep-ph/0610250}}.

\bibitem{Morisi:2009sc}
S.~Morisi and E.~Peinado, ``{An A4 model for lepton masses and mixings},'' {\em
  Phys. Rev.} {\bfseries D80} (2009) 113011,
\href{http://arxiv.org/abs/0910.4389}{{\ttfamily arXiv:0910.4389 [hep-ph]}}.

\bibitem{Beringer:1900zz}
{\bfseries Particle Data Group} Collaboration, J.~Beringer {\em et~al.},
  ``{Review of Particle Physics (RPP)},''
\href{http://dx.doi.org/10.1103/PhysRevD.86.010001}{{\em Phys.Rev.} {\bfseries
  D86} (2012) 010001}.

\bibitem{Ade:2013zuv}
{\bfseries Planck Collaboration} Collaboration, P.~Ade {\em et~al.}, ``{Planck
  2013 results. XVI. Cosmological parameters},''
\href{http://arxiv.org/abs/1303.5076}{{\ttfamily arXiv:1303.5076
  [astro-ph.CO]}}.

\bibitem{Bornschein:2003xi}
{\bfseries KATRIN Collaboration} Collaboration, L.~Bornschein, ``{KATRIN:
  Direct measurement of neutrino masses in the sub-Ev region},'' {\em eConf}
  {\bfseries C030626} (2003) FRAP14,
\href{http://arxiv.org/abs/hep-ex/0309007}{{\ttfamily arXiv:hep-ex/0309007
  [hep-ex]}}.


\end{thebibliography}

\providecommand{\href}[2]{#2}\begingroup\raggedright\endgroup

\end{document}